\documentclass[preprint,amsmath,amssymb]{revtex4}
\usepackage{graphicx}

 \def\Pom{{ I\!\!P}}
\def\Reg{{ I\!\!R}}
\def\gsim{\mathrel{\rlap{\lower4pt\hbox{\hskip1pt$\sim$}}
 \raise1pt\hbox{$>$}}}

 \newcommand\la{\langle}
 \newcommand\ra{\rangle}
 \newcommand\beq{\begin{equation}}
 \newcommand\noi{\noindent}
 \newcommand\eeq{\end{equation}}
 \newcommand\beqn{\begin{eqnarray}}
 \newcommand\eeqn{\end{eqnarray}}
\def\mb{\,\mbox{mb}}
\def\fm{\,\mbox{fm}}
\def\GeV{\,\mbox{GeV}}
\def\TeV{\,\mbox{TeV}}
\def\lsim{\mathrel{\rlap{\lower4pt\hbox{\hskip1pt$\sim$}}
    \raise1pt\hbox{$<$}}}         
\def\gsim{\mathrel{\rlap{\lower4pt\hbox{\hskip1pt$\sim$}}
    \raise1pt\hbox{$>$}}}         
\def\Re{\,{\rm Re}\,}
\def\Im{\,{\rm Im}\,}
\def\mb{\,\mbox{mb}}
\def\fm{\,\mbox{fm}}
\def\GeV{\,\mbox{GeV}}

\def\Journal#1#2#3#4{{#1}{\bf #2}, #3 (#4). }

\def\EPJA{{\it Eur. Phys. J.} {\bf A}}

\def\NPB{{\it Nucl. Phys.} {\bf B}}

\def\PLB{{\it Phys. Lett.} {\bf B}}

\def\PRL{\it Phys. Rev. Lett. }
\def\PREV{\it Phys. Rev. }
\def\PREP{\it Phys. Rep. }

\def\PRD{{\it Phys. Rev.} {\bf D}}
\def\PRC{{\it Phys. Rev.} {\bf C}}

\def\RMP{{\it Rev. Mod. Phys.} }

\def\JETP{\it JETP }
\def\JETPL{\it JETP Lett. }

\def\r{\vec r}

\def\Im{\mbox{Im}\,}
\def\Re{\mbox{Re}\,}
\def\mb{\,\mbox{mb}}
\def\fm{\,\mbox{fm}}
\def\GeV{\,\mbox{GeV}}

\def\TeV{\,\mbox{TeV}}
\def\eps{\varepsilon}

\def\Pom{\mathbb{P}}
\def\Reg{\mathbb{R}}
\def\lsim{\mathrel{\rlap{\lower4pt\hbox{\hskip1pt$\sim$}}
    \raise1pt\hbox{$<$}}}         
\def\gsim{\mathrel{\rlap{\lower4pt\hbox{\hskip1pt$\sim$}}
    \raise1pt\hbox{$>$}}}         

\def\halftext{.471\textwidth}
\def\st{\sigma_{tot}^{hN}}
\def\sel{\sigma_{el}^{hN}}

\def\sdd{\sigma_{sd}^{hN}}

\def\sta{\sigma_{tot}^{hA}}

\def\sq{\sigma_{\bar qq}}

\begin{document}

\vspace*{50px}
\title{Gribov inelastic shadowing\\ in the dipole representation\footnote{To be published in
"Gribov-85 Memorial Volume", World Scientific, 2016.}\vspace{4ex}}

\author{B. Z. Kopeliovich\vspace{4ex}}

\affiliation{\centerline{Departamento de F\'{\i}sica,
Universidad T\'ecnica Federico Santa Mar\'{\i}a; and}
Centro Cient\'ifico-Tecnol\'ogico de Valpara\'iso;
Casilla 110-V, Valpara\'iso, Chile
\vspace{8ex}}

\begin{abstract}
The dipole phenomenology, which has been quite successful applied to
various hard reactions, especially on nuclear targets, is applied for calculation of Gribov inelastic shadowing. This approach does not include ad hoc procedures, which are unavoidable
in calculations done in hadronic representation. Several examples of Gribov corrections
evaluated within the dipole description are presented.
\end{abstract}


\maketitle

\section{Introduction}

The Glauber model~\cite{glauber} was the first theoretical approach which  calculated the effects of shadowing in hadron-nucleus interactions. The model however, was essentially non-relativistic. In the pioneering papers by V.N.~Gribov~\cite{gribov69,gribov70}
it was realised that the length scales of interaction rising with energy significantly
change the pattern of hadron-nucleus interaction at high energies.
In particular, particles created in an inelastic collision with one nucleon, can be subsequently absorbed by another bound nucleons. Such corrections make the nuclear medium more transparent for hadrons, and consequently lead to a reduction of the total hadron-nucleus cross section. These corrections, called Gribov inelastic shadowing,
improve the Glauber model, making it well-founded.

\section{From the Glauber model to the Gribov-Glauber theory}

The amplitude of probability for a hadron to interact with the nucleus is one minus the probability amplitude of no interaction with any of the bound nucleons. So the $hA$ elastic amplitude at impact parameter $b$ has the eikonal 
form,
 \beq
 \Gamma^{hA}(\vec b;\{\vec s_j,z_j\}) =
1 - \prod_{k=1}^A\left[1-
 \Gamma^{hN}(\vec b-\vec s_k)\right]\ ,
 \label{1.40}
 \eeq
 where $\{\vec s_j,z_j\}$ denote the coordinates of the target nucleon
$N_j$. $i\Gamma^{hN}$ is the elastic scattering amplitude on a nucleon
normalized as,
 \beqn
\st &=& 2\int d^2b\,\Re\Gamma^{hN}(b);\nonumber\\
\sel&=& \int d^2b\, |\Gamma^{hN}(b)|^2\ .
\label{1.60}
 \eeqn
 
In the approximation of single particle nuclear density one can calculate
a matrix element between the nuclear ground states.
 \beqn
\left\la0\Bigl|\Gamma^{hA}(\vec b;\{\vec s_j,z_j\})
\Bigr|0\right\ra =
1-\left[1-{1\over A}\int d^2s\,
\Gamma^{hN}(s)\int\limits_{-\infty}^\infty dz\,
\rho_A(\vec b-\vec s,z)\right]^A\ ,
\label{1.80}
 \eeqn
 where 
 \beq
\rho_A(\vec b_1,z_1) = \int\prod_{i=2}^A
d^3r_i\,
|\Psi_A(\{\vec r_j\})|^2\ ,
\label{1.100}
 \eeq
 is the nuclear single particle density.
 
Eq.~(\ref{1.80}) is related via unitarity to the total $hA$ 
cross section,
 \beqn
\sta&=&2\Re\int d^2b\,\left\{1 -
\left[1-{1\over A}\int d^2s\,
\Gamma^{hN}(s)\,T_A(\vec b-\vec s)\right]^A\right\}
\nonumber\\ &\approx&
2\int d^2b\, \left\{1-
\exp\left[-{1\over2}\,\st\,(1-i\rho_{pp})\,
T^h_A(b)\right]\right\}\ ,
\label{1.120}
 \eeqn
 where $\rho_{pp}$ is the ratio of the real to imaginary parts
of the forward $pp$ elastic amplitude;
 \beq
 T^h_A(b)= \frac{2}{\st}\int d^2s\, 
\Re\Gamma^{hN}(s)\,T_A(\vec b-\vec s)\ ;
\label{1.140} 
 \eeq
 and
 \beq
T_A(b) = \int_{-\infty}^\infty dz\,\rho_A(b,z)\ ,
\label{1.160}
 \eeq
 is the nuclear thickness function. 
 We use Gaussian form of $\Gamma^{hN}(s)$ in what follows,
 \beq
\Re \Gamma^{hN}(s) =
\frac{\st}{4\pi B_{hN}}\,
\exp\left(\frac{-s^2}{2B_{hN}}\right)\ ,
\label{1.180}
 \eeq
 where $B_{hN}$ is the slope of the differential $hN$ elastic cross
section. Notice that the accuracy of the optical approximation (the second line in
(\ref{1.120})) is quite high for heavy nuclei. For the sake of simplicity, we use the optical form
throughout the paper although for numerical evaluations always rely on the accurate expression (the first line in (\ref{1.120})). 
The effective nuclear thickness, Eq.~(\ref{1.140}) implicitly contains energy dependence, which is extremely weak.

In what follows we also neglect the real part of the elastic amplitude, unless specified, since it gives a vanishing correction  $\sim 
\rho_{pp}^2/A^{2/3}$.

Besides the total cross section Eq.~(\ref{1.120}) one can calculate within the Glauber model also elastic, quasielastic (break-up of the nucleus without particle production) and inelastic cross section. One can find details of such calculations, as well as numerical results, in Ref.\cite{mine,xsect,kps-ciofi}, and below in Sect.~\ref{hA}.

The Glauber model is intensively used nowadays as a theoretical tool to study heavy ion collisions. However, this model is subject to significant Gribov corrections \cite{gribov69}.

\subsubsection{Intermediate state diffractive excitations}\label{sect-kk}
\label{sec:intermediate}

\begin{figure}[htb]
\centerline{\includegraphics[width= 8.5 cm]{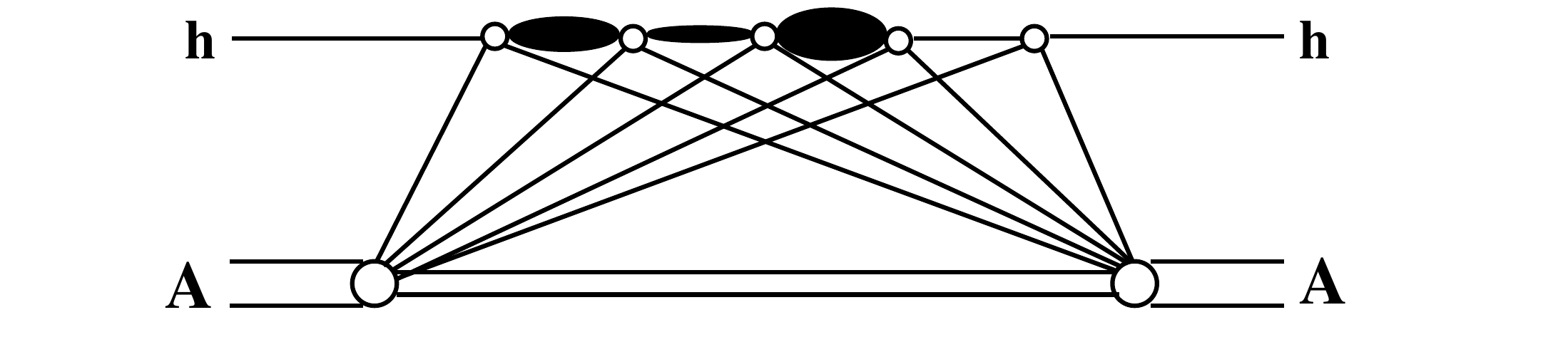}}
\caption{Diagonal and off-diagonal diffractive multiple excitations of the
projectile hadron in intermediate state}
\label{fig:diff}
\end{figure}

The Glauber model is a single-channel approximation, therefore it misses the
possibility of diffractive excitation of the projectile in the intermediate
state as is  illustrated in Fig.~\ref{fig:diff}.

Inclusion of multiple diffractive transitions between different excitations, like depicted in Fig.~\ref{fig:diff}, obviously is a challenge, because diffractive transitions between different excited states cannot be measured in diffractive processes. 
There is, however, one case free of these problems,
shadowing in hadron-deuteron interactions. In this case no interaction in the intermediate
state is possible, and knowledge of diffractive cross section $hN\to XN$
is sufficient for calculations of the inelastic correction with no further
assumptions. In this case the Gribov correction to the Glauber model for the total cross section has the simple form
\cite{gribov69},
 \beq
\Delta\sigma^{hd}_{tot} = - 
2\int dM^2\int dp_T^2\,
\frac{d\sdd}{dM^2dp_T^2}\,
F_d(4t).
\label{1.360}
 \eeq
Here $F(t)$ is the deuteron electromagnetic formfactor; 
$\sigma_{sd}^{hN}$ is the cross section of single diffractive 
dissociation $hN\to XN$ with longitudinal momentum transfer
 \beq
q_L=\frac{M^2-m_h^2}{2E_h};
\label{1.340}
 \eeq
and $t=-p_T^2-q_L^2$.

The formula for the lowest order inelastic
corrections to the total hadron-nucleus cross section was suggested in
\cite{kk},
 \beqn
\Delta\sigma^{hA}_{tot} &=& 
- 8\pi\int d^2b\,
e^{-{1\over2}\sigma^{hN}_{tot}T_A(b)}\!\!\!
\int\limits_{M_{min}^2}\!\! dM^2
\left.\frac{d\sigma_{sd}^{hN}}
{dM^2\,dp_T^2}\right|_{p_T=0}
\nonumber\\ &\times&
\int\limits_{-\infty}^{\infty}dz_1\,
\rho_A(b,z_1)
\int\limits_{z_1}^{\infty}dz_2\,
\rho_A(b,z_1)\,e^{iq_L(z_2-z_1)}\ ,
\label{1.320}
 \eeqn
 
\begin{figure}[htb]
\centerline{\includegraphics[width=7 cm]{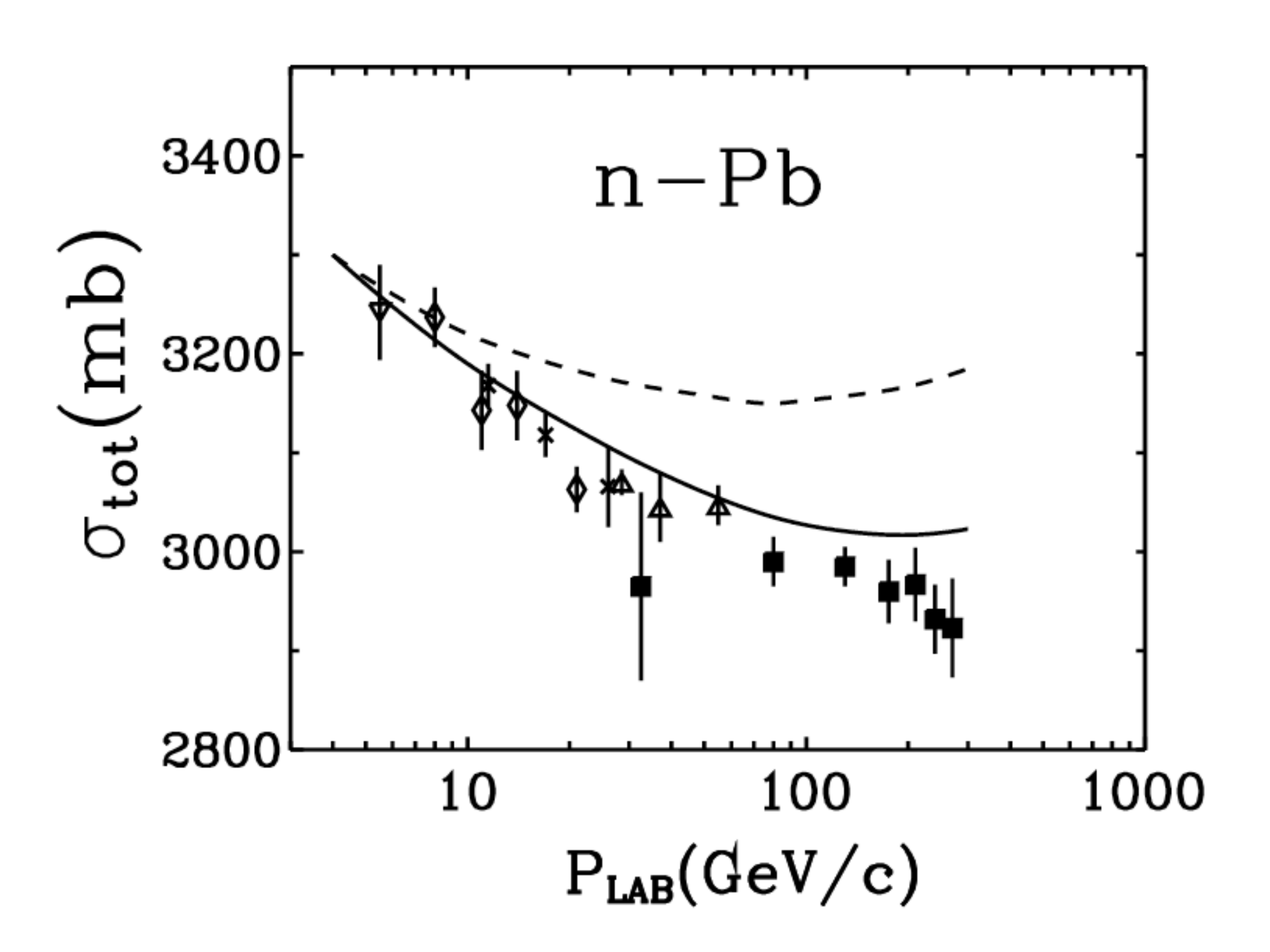}}
\caption{Data and calculations \cite{murthy} for the total neutron-lead
cross section. The dashed and solid curves correspond to the
Glauber model and  corrected for Gribov shadowing respectively.\label{fig:murthy}}
\end{figure}

This correction  takes care of the onset of
inelastic shadowing via phase shifts controlled by $q_L$ and does a good job describing data at low energies \cite{murthy,gsponer}, as one can also see
in Fig.~\ref{fig:murthy}. 
 
 In a similar way Gribov corrections can be calculated for neutrino-nucleus cross section
 at low $Q^2$, when PCAC hypothesis is at work.
 Then one can employ the Adler relation, which connects neutrino and pion induced
 cross sections at $Q^2=0$, and write the nucleus-to-proton ration of total neutrino-nucleus cross sections as \cite{nu-tot},
 \beqn
R^\nu_{A/N}&=&1 - \frac{8\pi}{A\sigma^{\nu N}_{tot}}
\left.\frac{d\sigma^{\nu\to\pi}_{diff}}{dp_T^2}\right|_{p_T=0}
\int d^2b\int\limits_{-\infty}^\infty dz_1\,\rho_A(b,z_1)
\nonumber\\&\times&
\int\limits_{-\infty}^{z_1} dz_2\,\rho_A(b,z_2)
e^{-iq_L^\pi(z_2-z_1)}\,e^{-{1\over2}\sigma^{\pi N}_{tot}T_A(b,z_2,z_1)},
\label{2.120}
\eeqn
where $q_L^\pi=(Q^2+m_{\pi}^2)/2\nu$.

At very low energy where $q_L^\pi$ is large, the second term in (\ref{2.120}) is suppressed, and the first one,  corresponding to the cross section proportional to $A$, dominates. At high energies $q_L^\pi\ll R_A$ can be neglected and the integrations over $z_{1,2}$ can be performed analytically \cite{nu-tot}. Due to the Adler relation the first term in (\ref{2.120}) and the volume part of the second term cancel, and the rest is the "surface" term $\propto A^{2/3}$,
\beq \left.\frac{d^2\sigma(\nu A\to
l\,X)}{dQ^2\,d\nu}\right|_{q_L\ll1/R_A} = 
\frac{G^2}{2\pi^2}\,f_\pi^2\,\frac{E-\nu}{E\nu}
\,\sigma(\pi A\to X), 
\label{2.40} 
\eeq 
as could be anticipated in accordance with the Adler relation. 
Notice that the high-energy regime actually starts at rather low energies if $Q^2\lsim
m_\pi^2$.  
 
\begin{figure}[htb]
\parbox{\halftext}{
\centerline{\includegraphics[width=5 cm]{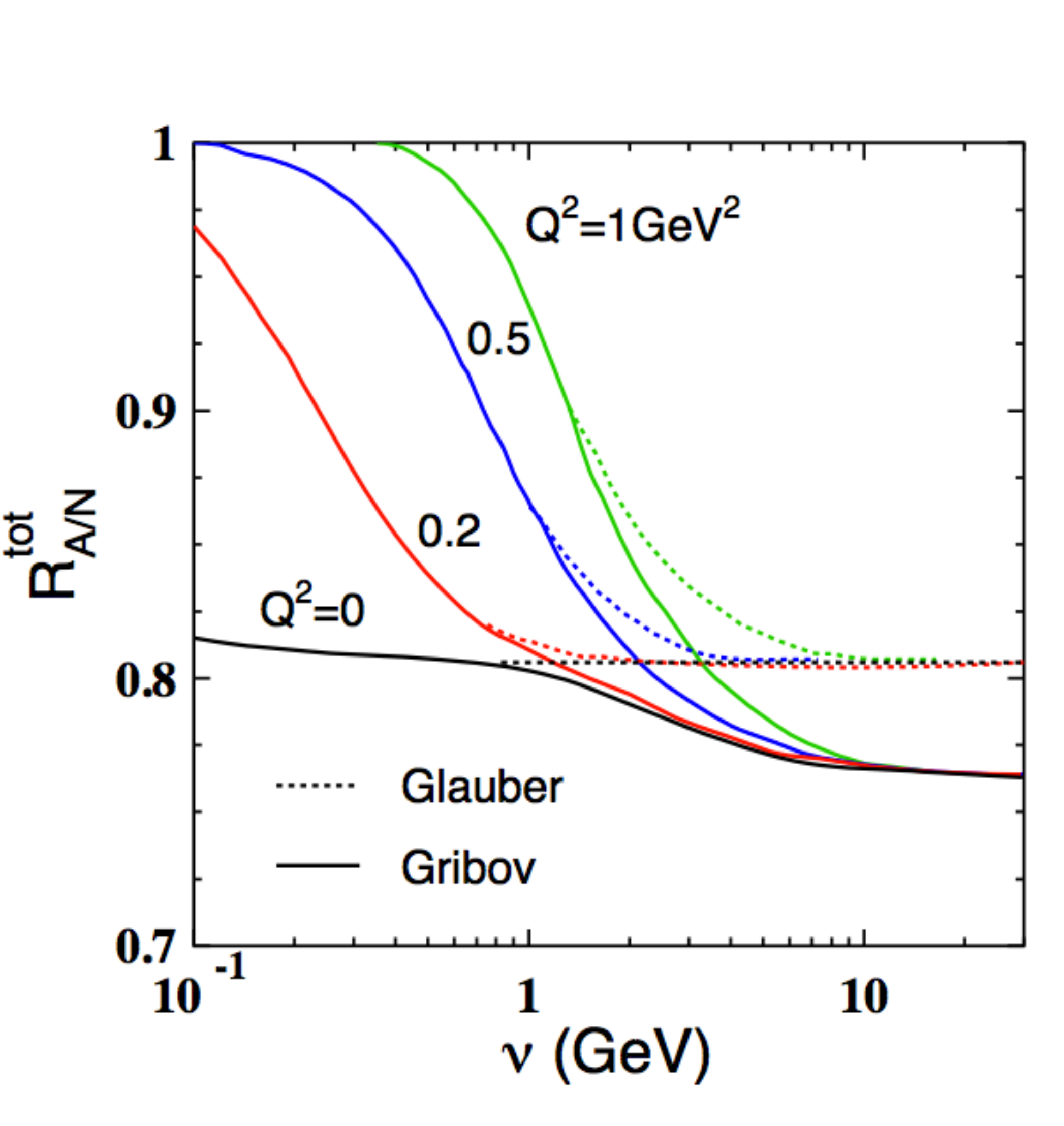}}
 \caption{The neon to nucleon ratio of total neutrino cross sections
 at different $Q^2$ \cite{nu-tot}. Dashed and solid curves
correspond to the Glauber and Gribov corrected calculations. }
 \label{fig:shad-1}}
\hfill
\parbox{\halftext}{
\centerline{\includegraphics[width=5 cm]{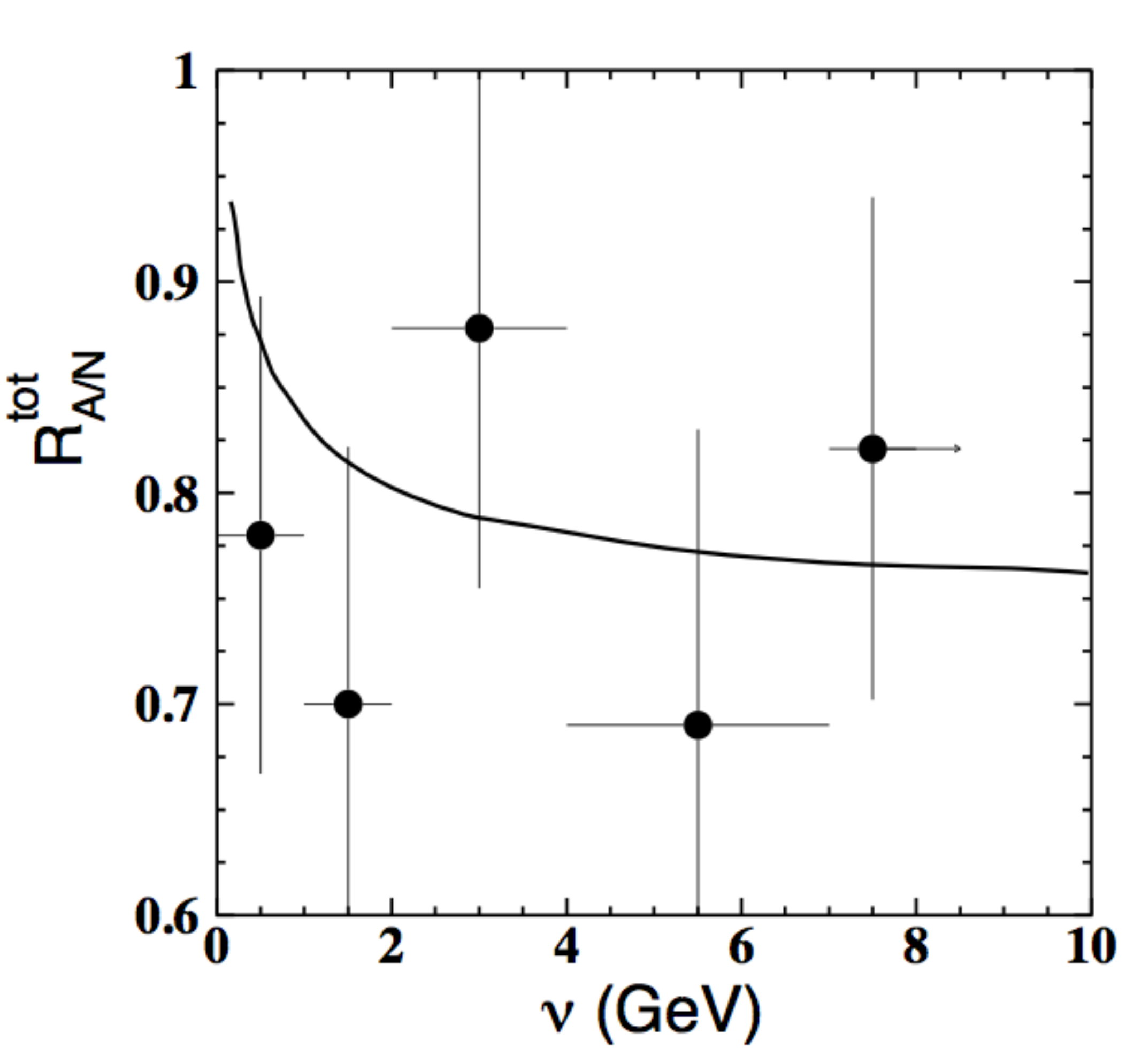}}
\caption{ The neon to proton ratio of the total neutrino cross sections,
calculated in Ref.\cite{nu-tot} for $x<0.2$ and $Q^2<0.2\GeV^2$. The data present 
the BEBC results \cite{wa59}.}\label{fig:shad-2}}
 \end{figure}

The results of numerical evaluation of the shadowing effect Eq.~(\ref{2.120})
for neon are plotted in Fig.~\ref{fig:shad-1}.
The calculations \cite{nu-tot} done within 
the Glauber approximation, Eq.~(\ref{2.120}), and also including the Gribov's inelastic
corrections (important at high energies $\nu>m_{a_1}^2R_A$) are plotted in Fig.~\ref{fig:shad-1} by solid curves as function of energy, for different $Q^2$. 

The calculated shadowing effects are compared with BEBC data \cite{wa59} in Fig.~\ref{fig:shad-2}.
 As was anticipated, the shadowing exposes an early onset, and a significant suppression occurs at small $Q^2$ in
the low energy range of hundreds MeV. This is an outstanding feature of
the axial current. This seems to be supported by data, although 
 with a rather poor statistics.

 Concluding this section, we notice that although the approximation of lowest  order Grobov corrections does in some cases good job,  the
higher order off-diagonal transitions neglected in (\ref{1.320}) and (\ref{2.120}),  might be important, but unknown. Indeed, the
intermediate state $X$ has definite mass $M$, but no definite
cross section. It was fixed in (\ref{1.320}) at $\sigma_{tot}^{hN}$ with no justification.

Thus, the lowest order Gribov correction Eq.~(\ref{1.320}) has a nice feature
of including phase shifts, which allows to describe the onset of Gribov shadowing,
like is presented in Fig.~\ref{fig:murthy}. However, the uncertainties related to the missed higher order corrections and the unknown absorption in (\ref{1.320}) seem to be incurable, while one works in the hadronic (eigenstates of the mass matrix) representation.

\subsection{Interaction eigenstate representation}\label{eigen}

If a hadron were an eigenstate of interaction, i.e. could undergo only
elastic scattering (as a shadow of inelastic channels) and no diffractive
excitation was possible, the Glauber formula would be exact and no
inelastic shadowing corrections were needed. This simple observation
gives a hint that one should switch from the hadronic basis to a complete set of mutually orthogonal eigenstates of the scattering amplitude operator. This
was the driving idea of the description of diffraction in terms of elastic
amplitudes \cite{pom,gw}, and becomes a powerful tool for calculation of
inelastic shadowing corrections in all orders of multiple interactions
\cite{kl,zkl}. Notice that this idea also was used by Gribov \cite{gribov69} to explain
the role of the coherence length increasing with energy, on the example of an incoming proton, fluctuating to a nucleon-pion pair.

Physical states (including leptons, photons) can be expanded
over the complete set of states $|k\ra$,
 \beq
|h\ra=\sum\limits_{k}\,\Psi^h_k\,|k\ra\ , 
\label{b.1}
 \eeq 
 which are the eigenstates of the scattering amplitude operator, $\hat f\,|k\ra=
 f_{el}^{kN}|k\ra$. For the sake of simplicity we neglect the real part of the amplitude, i.e. assume that $f_{el}^{kN}=i\,\sigma_{tot}^{kN}/2$.
 
$\Psi^h_k$ in (\ref{b.1}) are weight factors (amplitudes) of the Fock state
decomposition. They obey the orthogonality conditions,
 \beqn
\sum\limits_{k}\,\left(\Psi^{h'}_k\right)^{\dagger}\,\Psi^h_k
&=&\delta_{h\,h'}\ ;
\nonumber\\
\sum\limits_{h}\,\left(\Psi^{h}_l\right)^{\dagger}\,\Psi^h_k
&=&\delta_{lk}\ .
\label{b.2}
 \eeqn

We also assume
that the amplitude is integrated over impact parameter, {\it i.e.} that
the forward elastic amplitude is normalized as
$|f_{el}^{kN}|^2=4\,\pi\,d\sigma_{el}^{kN}/dt|_{t=0}$. 
The amplitudes of elastic $f_{el}(hh)$ and off diagonal diffractive
$f_{sd}(hh')$ transitions can be expressed as,
 \beq
f_{el}^{hN}=2i\,\sum\limits_k\,\left|
\Psi^h_k\right|^2\,\sigma_{tot}^{kN}
\equiv 2i\,\la\sigma\ra\ ;
\label{b.3}
 \eeq
 \beq
f_{sd}^{hN}(h\to h')=
2i\,\sum\limits_k\, (\Psi^{h'}_k)^{\dagger}\,
\Psi^h_k\,\sigma_{tot}^{kN}\ . 
\label{b.4}
 \eeq
 Note that if all the eigenamplitudes were equal, the diffractive
amplitude (\ref{b.4}) would vanish due to the orthogonality relation,
(\ref{b.2}). The physical reason is obvious. If all the $f_{el}^{kN}$ are
equal, the interaction does not affect the coherence between 
different eigen components $|k\ra$ of the projectile hadron $|h\ra$.
Therefore, the off-diagonal transitions are possible only due to differences
between the eigenamplitudes.

By summing up in the diffractive cross section
using completeness  Eq.~(\ref{b.2}), and
excluding the elastic channels, one gets \cite{kl,mp,zkl},
 \beqn
16\pi\,\frac{d\sigma^{hN}_{sd}}{dt}\biggr|_{t=0}&=&
\sum\limits_k \left|\Psi^h_k\right|^2
\left(\sigma^{kN}_{tot}\right)^2
- \biggl(\sum\limits_i 
\left|\Psi^h_k\right|^2\sigma^{kN}_{tot}\biggr)^2
\nonumber\\
&\equiv& \left\la\left(\sigma^{kN}_{tot}\right)^2\right\ra - 
\left\la\sigma^{kN}_{tot}\right\ra^2\ .
\label{b.6}
 \eeqn

Each of the eigenstates propagating through the nucleus can experience only elastic
scatterings, so the Glauber eikonal approximation becomes
exact for such a state.  Then, the cross sections for
hadron-nucleus collisions should be averaged over the Foch states, chosen to be interaction eigenstates \cite{kl,zkl},
 \beq
\sigma^{hA}_{tot} = 2\int d^2b\,\left\{1 -
\left\la\exp\left[-{1\over2}\,\sigma^{kN}_{tot}\,T^h_A(b)\right]
\right\ra\right\}.
\label{b.10}
 \eeq

This is to be compared with the Glauber approximation. The difference is
obvious, in Eq.~(\ref{b.10}) the exponential is averaged,
while in the Glauber approximation the exponent is averaged,
 \beq
\sigma^{hA}_{tot}\Bigr|_{Gl} = 2\int d^2b\,\left\{1 -
\exp\left[-{1\over2}\,\left\la\sigma_{tot}^{kN}\right\ra\, T^h_A(b)\right]
\right\},
\label{b.30}
 \eeq
 where $\la\sigma^{kN}_{tot}\ra = \st$. The difference between
Eqs.~(\ref{b.10}) and (\ref{b.30}), is the Gribov inelastic correction
calculated in all orders of opacity expansion, which was impossible to do within hadronic representation (see above). This result can be compared with the expression (\ref{1.320})
for the lowest order correction expanding the exponentials in (\ref{b.10}) and
(\ref{b.30}) in number of collisions up to the lowest order.
Applying (\ref{b.6}) we find,
 \beqn
\sigma^{hA}_{tot} - \sigma^{hA}_{tot}\Bigr|_{Gl} &=&
\int d^2b\, {1\over4}\,\Bigl[\left\la\sigma^{kN}_{tot}\right\ra^2 -
\left\la\left(\sigma^{kN}_{tot}\right)^2\right\ra\Bigr]\,T^h_A(b)^2
\nonumber\\ &=&
- 4\pi\int d^2b\,T^h_A(b)^2\int dM^2\,
\frac{d\sigma^h_{sd}}{dM^2dt}\biggr|_{t=0}.
 \eeqn
 This result is indeed identical to Eq.~(\ref{1.320}), if to neglect the 
phase shift vanishing at high energies, and also to expand the 
exponential.

\section{Color dipoles as eigenstates of interactions}

The dipole representation in QCD, first proposed in Ref.\cite{zkl}, 
allows to calculated Gribov corrections in all orders, because a high energy dipole
is an eigenstate of interaction. Indeed, the transverse dipole separation, which controls the scattering amplitude, is preserved during propagation and remains intact after multiple soft interactions within
a nuclear target.

To perform proper averaging in (\ref{b.10}) one should expend the beam particle
(hadrons, vector or axial currents) over Fock states. Each Fock component is a colorless dipole, consisted of two or more partons. After averaging over intrinsic dipole distances, one should sum up different Fock states with proper weights. 

\subsection{Deep-inelastic scattering}

The parton model interpretation of the space-time development of reactions is not Lorentz invariant. While Deep-inelastic scattering (DIS) at small $x\ll1$ is interpreted in the Bjorken reference frame as absorption of the virtual photon by a target parton carrying corresponding fractional momentum $x$ of the target, in the target rest frame it looks differently.
The high energy virtual photon fluctuates into a $\bar qq$ pair, which interacts with the targets and is produced on mass shell. 

\begin{figure}
\centerline{\includegraphics[width=8 cm]{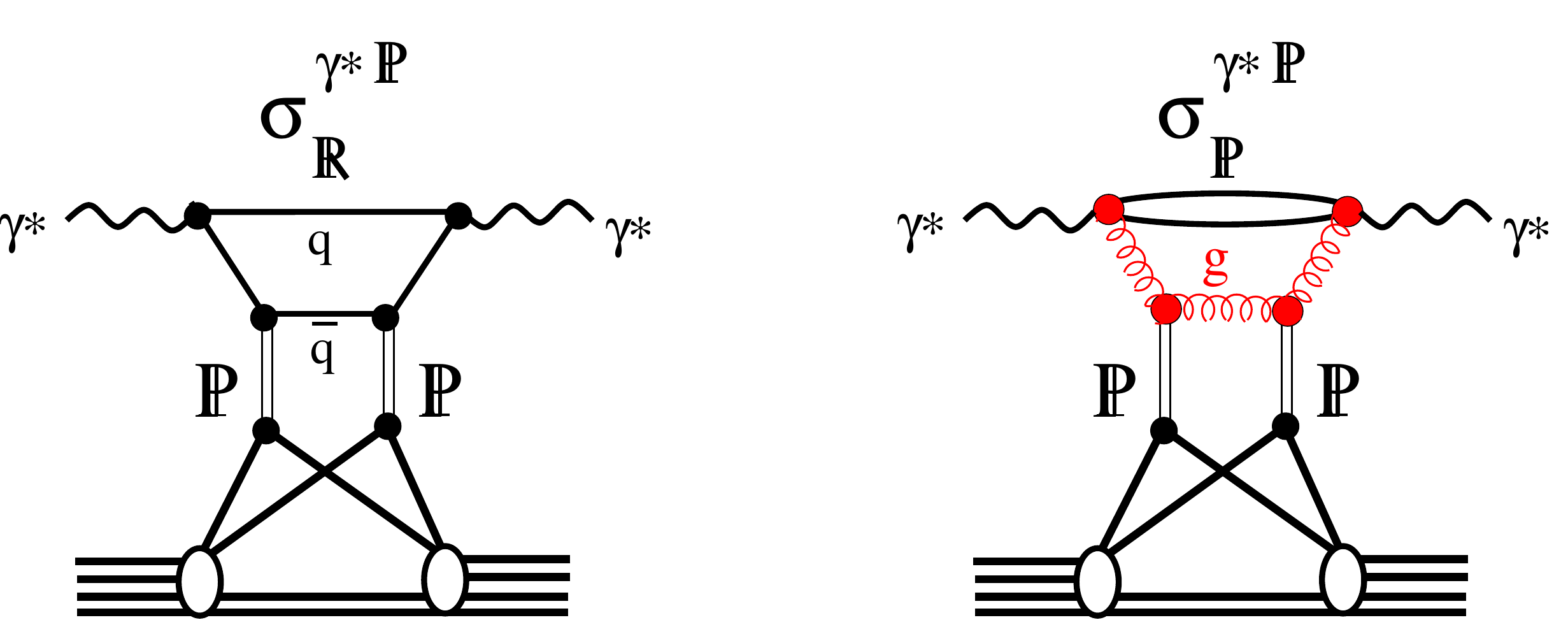}} 
\caption{Gribov corrections with $\bar qq$ and $\bar qq+g$ intermediate states. }
\label{fig:pom-reg}
 \end{figure}

The two main contributions to the diffractive cross section in 
(\ref{1.320}) at high energies come from the triple-Regge terms $\Pom\Pom\Reg$ 
and $\Pom\Pom\Pom$, which generate $1/M^3$ and $1/M^2$ mass dependences respectively. The quark-gluon structure of these contributions in diffraction is illustrated in Fig.~\ref{fig:pom-reg}.
Apparently, the $\bar qq$ and $\bar qqg$ states, propagating through the nucleus,
attenuate with an absorptive cross section, which cannot be trivially fixed, like was done
in (\ref{1.320}), but should be properly treated within the dipole approach.

\subsubsection{"Frozen" dipoles}

If the lifetime of partonic fluctuations of a photon significantly exceeds the nuclear size, the dipole approach allows to calculate easily the shadowing effects in DIS. Indeed, in this case one can rely on the ``frozen" approximation Eq.~(\ref{b.10}), which for interaction of a virtual photon has the form,
\beqn
\sigma_{T,L}^{\gamma^*A}(x,Q^2)&=&
2\int d^2b\int\limits_0^1d\alpha\int d^2r_T
\left|\Psi_{q\bar q}^{T,L}(\alpha,r_T,Q^2)\right|^2
\nonumber\\ &\times&
\left[1-e^{-{1\over2}\sigma_{q\bar q}(r_T,x)T_A(b)}\right],
\label{4.200}
\eeqn
where $r_T$ is the dipole transverse separation; $\alpha$ is the fractional light-cone momentum of the quark;
the distribution functions $\Psi_{q\bar q}^{T,L}(\alpha,r_T)$ in the quadratic form read,
\beqn
\left|\Psi_{q\bar q}^{T}(\alpha,r_T)\right|^2&=&
\frac{2N_c\alpha_{em}}{(2\pi)^2}\sum\limits_{f=1}^{N_f}Z_f^2
\left\{\left[1-2\alpha(1-\alpha)\right]\eps^2{\rm K}^2_1(\eps r_T)
\right.\nonumber\\ &+&\left.
m_f^2{\rm K}^2_0(\eps r_T)\right\};
\label{4.210}\\
\label{psil}
\left|\Psi_{q\bar q}^{L}(\alpha,r_T)\right|^2&=&
\frac{8N_c\alpha_{em}}{(2\pi)^2}\sum\limits_{f=1}^{N_f}Z_f^2
Q^2\alpha^2(1-\alpha)^2{\rm K}^2_0(\eps r_T).
\label{4.220}
\eeqn

The advantage of the dipole description is pretty obvious,
Eq.~(\ref{4.200}) includes Gribov inelastic shadowing corrections to all orders of multiple interactions~\cite{zkl},
what is hardly possible within hadronic representation.

On the other hand, the dipoles having a definite size, do not have any definite mass, therefore the phase shifts between amplitudes on different nucleons cannot be calculated as simple as in Eq.~(\ref{1.320}).  A solution for this problem was proposed in Ref.\cite{krt1,zakharov}.

\subsubsection{Path integral technique}

The lifetime of the "frozen" dipole, or coherence length (or time), is given by,
\beq
l_c=\frac{2\nu}{Q^2+M_{\bar qq}^2},
\label{bh100}
\eeq
where $\nu$ is the dipole energy in the target rest frame; $M_{\bar qq}^2=(m_q^2+k_T^2)/\alpha(1-\alpha)$. At very small Bjorken
$x=Q^2/2m_N\nu\ll1$ the coherence length can significantly exceed the nuclear size, $l_c\gg R_A$, and one can safely rely on the "frozen" approximation.
However, if $t_c\lsim R_A$ such an approximation is not appropriate and one should correct for the dipole size fluctuations during propagation through the nucleus,
which corresponds to inclusion of the phase shifts between DIS amplitudes on different bound nucleons in Eq.~(\ref{1.320}). Within the dipole description this can be done employing the path integral technique \cite{feynman}, which sums up different propagation paths of the partons.

\begin{figure}[htb]
\centerline{\includegraphics[width=6 cm]{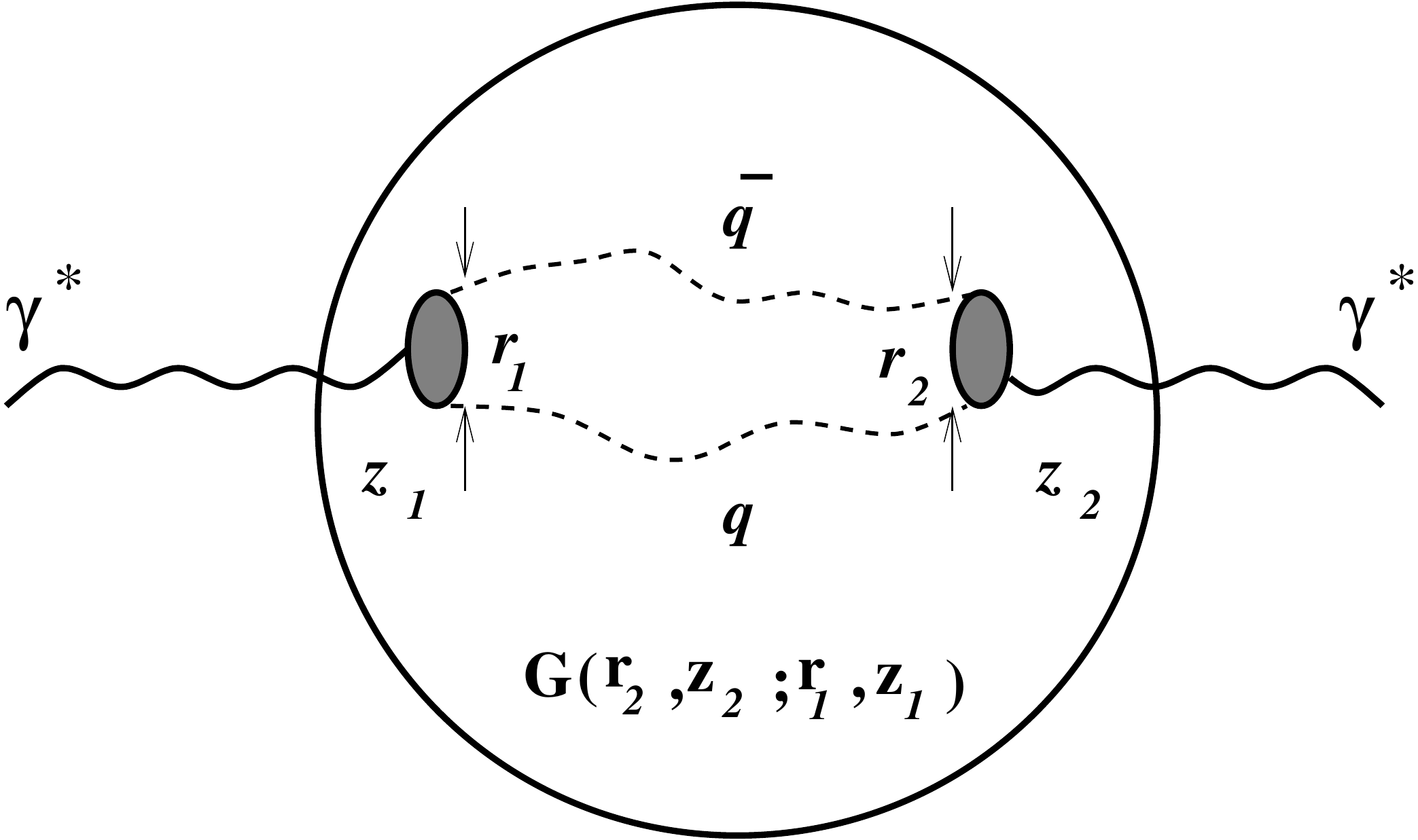}}
 \caption{Propagation of a $q\bar q$-pair through a nucleus between points with longitudinal coordinates $z_1$ and $z_2$.  The evolution of the $\bar qq$ separation from the initial, $\r_1$,  up to the final, $\r_2$, due to  the transverse motion of the quarks, is described by the
Green's function $G\left(\vec\rho_2,z_2;\vec\rho_1,z_1\right)$, a solution of Eq.~(\ref{4.260}). }
 \label{fig:GF}
\end{figure}

For a $\bar qq$ component of a transversely or longitudinally polarized virtual photon Eq.~(\ref{4.200}) should be replaced by,
\beqn
\Bigl(\sigma^{\gamma^*A}_{tot}\Bigr)^{T,L}\! &=& 
A\Bigl(\sigma^{\gamma^*N}_{tot}\Bigr)^{T,L}
\nonumber\\
&-& \frac{1}{2} Re\int d^2b
\!\int\limits_0^1 d\alpha
\!\int\limits_{-\infty}^{\infty} dz_1 \!\int\limits_{z_1}^{\infty} dz_2
\!\int d^2r_1\int d^2r_2
\nonumber\\ & \times &
\Psi^{T,L^\dagger}_{\bar qq}\!\!\left(\varepsilon,\r_2\right)
\rho_A\left(b,z_2\right)\sigma_{q\bar{q}}^N\left(s,\r_2\right)
G\left(\vec r_2,z_2\,|\,\vec r_1,z_1\right)
\nonumber\\ & \times &
\rho_A\left(b,z_1\right)\sigma_{q\bar{q}}^N\left(s,\r_1\right)
\Psi^{T,L}_{\bar qq}\!\left(\varepsilon,\r_1\right).
\label{4.240}
\eeqn
The Green's function
$G\left(\vec r_2,z_2\,|\,\vec r_1,z_1\right)$ describes propagation 
of a  $\bar qq$ pair  in an
absorptive medium, having initial separation $\r_1$ at the initial position $z_1$, up to the point $z_2$, where it gets separation $\r_2$, as is illustrated in Fig.~\ref{fig:GF}. 

It satisfies the evolution equation \cite{kz91,kst1,zakharov},
\beq
\left[i\frac{\partial}{\partial z_2}
+\frac{\Delta_\perp\left(r_2\right)-\varepsilon^2}
{2\nu\alpha\left(1-\alpha\right)}
+U(r_2,z_2)
\right]
G\left(\vec r_2,z_2\,|\,\vec r_1,z_1\right)
= 0
\label{4.260}
\eeq
The light-cone potential in the left-hand side of
this equation describes nonperturbative interactions within the dipole, and its absorption in the medium.
The real part the potential responsible for nonperturbative quark interactions was modelled and fitted to data of $F_2^p$ in Ref.\cite{kst2}. Here we fix $\Re U(r_2,z_2)=0$, and treat quarks as free particles for the sake of simplicity.
The imaginary part of the potential describes the attenuation of the dipole in the medium,
\beq
\Im U(r,z)=-{1\over2}\,\sigma_{\bar qq}(r)\,\rho_A(b,z).
\label{4.270}
\eeq

\begin{figure}[htb]
\centerline{\includegraphics[width=8cm]{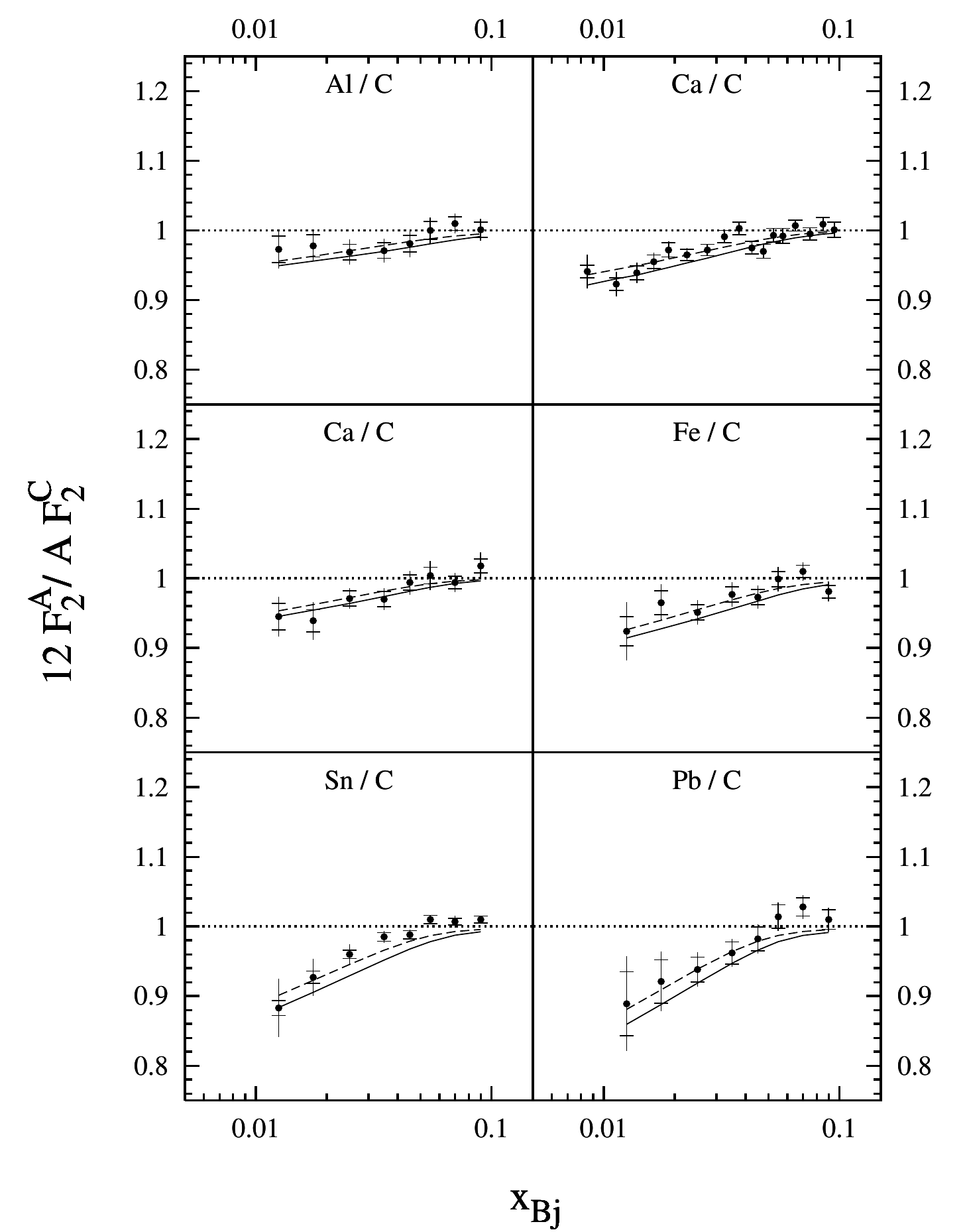}}
 \caption{Comparison between calculations for shadowing in DIS and experimental
      data from NMC \cite{nmc,nmc951}
      for the structure functions of
      different nuclei relative to carbon.  
$Q^2$ ranges within $3\le Q^2\le 17\GeV^2$. 
The solid and dashed curves are calculated including or 
excluding the real part of the potential respectively.
}
 \label{fig:nmcc}
\end{figure}

The numerical results of the calculations, which are performed either disregarding or including the real part of the potential, modelled in Ref.\cite{kst2},
are plotted in Fig.~\ref{fig:nmcc} by dashed and solid curves respectively. One can see that inclusion of the nonperturbative effects does not lead to a significant change of the magnitude of shadowing. Comparison with NMC data \cite{nmc,nmc951} shows pretty good agreement. 
We remind that this is a parameter-free description. More example of comparison with data can be found in Ref.\cite{krt2}. Notice also that we
did not include any mechanism of nuclear enhancement at $x>0.1$, the effect
called antishadowing, which affect the small-$x$ region as well.

\subsubsection{Gluon shadowing}

As was mentioned above, the contribution of all Fock components in the cross section should be summed up
\beq
\sigma_{tot}^{\gamma^*A} = A\,\sigma_{tot}^{\gamma^*N}\,
-\, \Delta\sigma_{tot}(\bar qq)\,
-\, \Delta\sigma_{tot}(\bar qqg)\, 
-\, \Delta\sigma_{tot}(\bar qq2g)\, -\,...
\label{4.1}
\eeq
The next after the $\bar qq$ Fock component is $\bar qq+g$.
It has a considerably shorter coherence time, compared with $\bar qq$, because
of specific nonperturbative effects increasing the mean transverse momentum of gluons
\cite{kst2,spots}. Correspondingly, the next component $|\bar qq2g\ra$ has even a much shorter coherence time $t_c$, which makes the 4th term in (\ref{4.1}) negligibly small within the currently achieved kinematic range.  Thus, we keep only first two low Fock components.

Notice that a successful attempt to sum up all Fock components was done in the form of Balitsky-Kovchegov equation \cite{b,k}

\begin{figure}
\centerline{\includegraphics[width=6 cm]{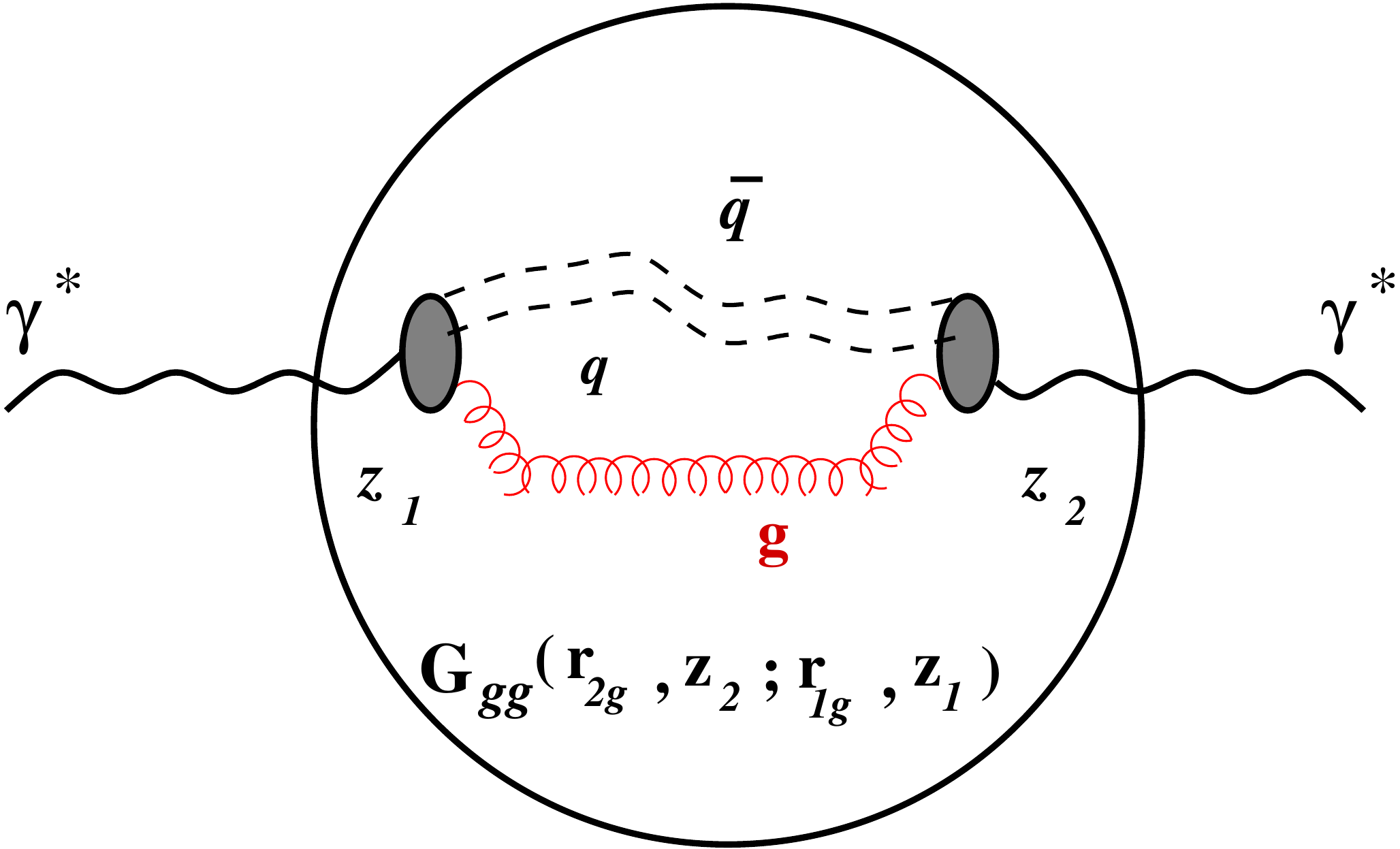}}
\caption{Propagation through a nucleus of the $q\bar q-g$ fluctuation of a longitudinally polarized photon.  
Neglecting the small, $\sim1/Q^2$ size of the color-octet $\bar qq$ pair, the effective octet-octet dipole propagation is described by the Green's function $G_{gg}(\r_{2g},z_2;\r_{1g},z_1)$. 
}
\label{fig:GF-g}
 \end{figure}

The 3rd term in the total cross section Eq.~(\ref{4.1}) is illustrated in Fig.~\ref{fig:GF-g}.

Differently from the case of a $|\bar qq\ra$ Fock state,
where we found that  at high $Q^2$ perturbative QCD can be safely used for shadowing calculations,
the nonperturbative effects remain important for the
$|\bar q\,q\,g\ra$ component even for highly virtual photons.
High $Q^2$ squeezes the $\bar qq$ pair down to a size $\sim 1/Q$,
while the mean quark-gluon separation at $\alpha_g \ll 1$ 
depends on the strength of nonperturbative gluon interaction which
is characterised in this limit by a small  separation $r_0\approx 0.3\fm$ \cite{kst2}.
which considerably smaller than the confinement radius $1/\Lambda_{QCD}$. This is confirmed
by various experimental observations \cite{spots}, in particular by the observed strong suppression of the 
diffractive gluon radiation \cite{kst2}. In nonperturbative QCD models this scale is related to the instanton size \cite{shuryak,zahed}.

The nonperturbative quark-gluon wave function 
was found in Ref.\cite{kst2} to have the form,
\beq
\Psi_{qg}\left(\vec r_{g},\alpha_g\right)\Bigr|_{\alpha_g\ll1}=
- \frac{2i}{\pi}\,
\sqrt\frac{\alpha_s}{3}\ 
\frac{\vec e\cdot\vec r_{g}}{r^2_{g}}\,
\exp\left(-\frac{r^2_{g}}{2r_0^2}\right)\ ,
\label{4.460}
\eeq
where $\vec e$ is the gluon polarization vector.

For $Q^2\gg 1/r_0^2$ the $\bar qq$ is small,
$r^2_{\bar qq} \ll r_g^2$, and one can treat the $\bar qqg$ system
as a color octet-octet dipole, as is illustrated in Fig.~\ref{fig:GF-g}.
Then the three-body Green's function factorizes,
\beq
G_{q\bar qg}\left(\vec r_{2g},
\vec r_{2},z_2;\vec r_{1g},\vec r_{1},z_1
\right)\Rightarrow
G_{q\bar q}\left(\vec r_{2},z_2;
\vec r_{1},z_1\right)
G_{gg}\left(\vec r_{2g},z_2;\vec r_{1g},z_1\right).
\label{4.400}
\eeq
The color octet-octet  Green function $G_{gg}$, describing the propagation of a glue-glue dipole with $\alpha_g\ll1$ through the medium, satisfies the simplified  evolution equation \cite{kst2},
\beq
\left[i\frac{\partial}{\partial z_2}-\frac{Q^2}{2\nu}
-V(\vec r_{2g},z_2)
\right]
G_{gg}\left(\vec r_{2g},z_2;\vec r_{1g},z_1\right)
=0
\label{4.420}
\eeq
Here 
\beq
\Im V(\vec r_{2g},z)=-{1\over2}\sigma_{gg}(r,x)\rho_A(z),
\label{bh200}
\eeq
with the color-octet dipole cross section, which reads,
\beq
\sigma_{gg}\left(r,x\right)
=\frac{9}{4}\,
\sigma_{q\bar q}\left(r,x\right).
\label{4.480}
\eeq

The real part of the potential must  correctly reproduce the wave function Eq.~(\ref{4.460}).
\beq
\Re V(\vec r_{2g},z)=
\frac{r_{2g}^2}
{2\nu \alpha_gr_0^4}
\label{bh300}
\eeq

Longitudinal photons can be used to disentangle between the effects of higher twist quark shadowing (2d term in (\ref{4.1})), and leading twist gluon shadowing (3rd term in (\ref{4.1})). The contribution, which mixes up these two types of Gribov corrections, come from so called aligned jet configurations \cite{bjorken} of the $\bar qq$ pair.
Namely the mean $\bar qq$ separation $r_{\bar qq}^2\sim Q^2/\alpha_q(1-\alpha_q)$
is small, unless the large value of $Q^2$ is compensated by smallness of $\alpha_q$ or $(1-\alpha_q)$. In the wave function of longitudinal photons, Eq.~(\ref{4.220}) such aligned-jet configurations are suppressed, so shadowing of longitudinal photons 
should represent the net effect of gluon shadowing.

While the distance between the $q$ and the $\bar q$ is small, of order $1/Q^2$,  the
gluon can propagate relatively far at a distance $r_g\sim r_0$  from the $q\bar q$-pair, which after the emission of the gluon is in a color-octet state. Therefore, the
entire $|q\bar qg\ra$-system appears as a octet-octet $gg$-dipole, and the shadowing
correction to the longitudinal cross section directly gives the magnitude of gluon 
shadowing, which we
want to calculate.

 Thus, the cross section of longitudinal photons is proportional to the gluon distribution function, therefore,

\beq
\frac{g_A(x,Q^2)}{g_N(x,Q^2)}\approx
\frac{\sigma^L_A(x,Q^2)}{\sigma^L_N(x,Q^2)}
\label{4.520}
\eeq

The shadowing 
correction to $\sigma^L_A(x,Q^2)$ has the form (compare with (\ref{4.240})),
\beqn
\Delta\sigma^L_A(x,Q^2) &=&- \Re\int d^2b
\int\limits_{-\infty}^{\infty} dz_1
\int\limits_{z_1}^{\infty} dz_2\,
\rho_A(b,z_1)\rho_A(b,z_2)
\nonumber\\ &\times&
\int d^2r_{2g}\,d^2r_{2\bar qq}\,d^2r_{1g}\,d^2r_{1\bar qq}
\int d\alpha_q\,d{\rm ln}(\alpha_g)
F^{\dagger}_{\gamma^*\to\bar qqg}
(\vec r_{2g},\vec r_{2\bar qq},\alpha_q,\alpha_g)
\nonumber\\ &\times& 
G_{\bar qqg}(\vec r_{2g},\vec r_{2\bar qq},z_2;\vec r_{1g},\vec r_{1\bar qq},z_1)
F_{\gamma^*\to\bar qqg}(\vec r_{1g},\vec r_{1\bar qq},\alpha_q,\alpha_g)
\label{4.540}
\eeqn
Here
\beq
F^{\dagger}_{\gamma^*\to\bar qqg}
(\vec r_{g},\vec r_{\bar qq},\alpha_q,\alpha_g)
=
-\,\Psi^L_{\bar qq}\left(\vec r_{\bar qq},
\alpha_q\right)\ \vec r_{g}\cdot\vec\nabla\,
\Psi_{qg}\left(\vec r_{g}\right)\ 
\sigma_{gg}^N\left(x,r_{g}\right)\ ,
\label{4.440}
\eeq

Assuming $Q^2\gg 1/r_0^2$ we can neglect $r_{\bar qq}\ll r_g$. The net diffractive amplitude $F_{\gamma^*\to\bar qqg}(\vec r_{1g},\vec r_{1\bar qq},\alpha_q,\alpha_g)$ takes the form of Eq.~(\ref{4.440}), and we can rely on the factorized relation (\ref{4.400}) for the 3-body Green's function, with equation (\ref{4.420}) for the evolution of the gluonic dipole. 

\begin{figure}[htb]
\centerline{\includegraphics[width=7 cm]{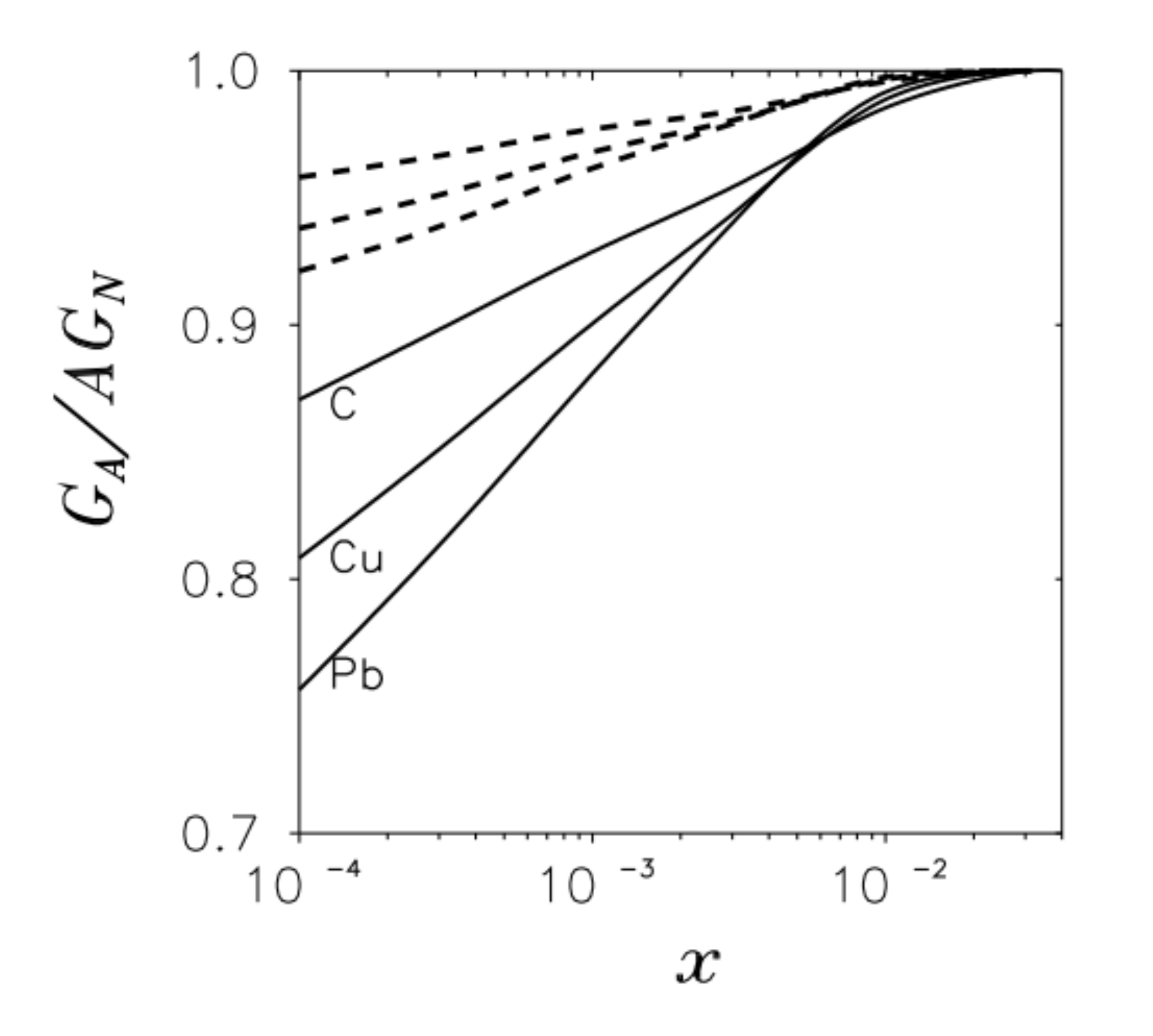}}
 \caption{Ratio (\ref{4.600}) for carbon, copper and lead
at small Bjorken $x$ and $Q^2 = 4\GeV^2$ (solid curves) and $40\GeV^2$ (dashed curves).
 }
 \label{fig:glue}
\end{figure}

The results of numerical calculation of (\ref{4.540}) for the ratio
\beq
R_g(x,Q^2)=\frac{g_A(x,Q^2)}{A\,g_N(x,Q^2)},
\label{4.600}
\eeq
are depicted in Fig.~\ref{fig:glue} as function of Bjorken $x$ for $Q^2=4$ and $40\GeV^2$.
The predicted small magnitude of gluon shadowing was confirmed by next-to-leading
(NLO) global analyses of DIS data \cite{florian,kumano}. It also goes along with the
well known smallness of the triple-Pomeron coupling, measured at low virtuality
\cite{kklp}. The latter controls large mass diffraction,
which proceeds via gluon radiation, and its smallness leads to suppression of gluon radiation and gluon shadowing related to corresponding Gribov corrections.
The observed weakness of these effects is interpreted in the dipole approach a smallness
of the parameter $r_0\approx0.3\fm$ in Eq.~(\ref{4.460}) \cite{kst2,spots}.

\subsection{Hadron-nucleus cross sections}\label{hA}

Applications of the dipole approach to calculation of Gribov corrections to the hadron-nucleus total cross sections contains more uncertainties and modelling compared with
hard reactions, like DIS. Nevertheless, it allows to make a progress compared with the hadronic representation, which involves ad hoc assumptions about the interaction cross section, of an excited hadronic state, and the unknown higher terms in the opacity 
expansion. 

\subsubsection{Excitation of the valence quark skeleton of the 
proton}

First of all one should rely on a parametrization of the dipole cross section,
which allows an extension to the soft, large separation region. Following
\cite{kst2,xsect} we chose the saturated shape of the cross section, which rises as $r_T^2$
at small $r_T$, but levels off at large $r_T$,
 \beq
\sigma_{\bar qq}(r_T,s)=\sigma_0(s)\,\left[
1-{\rm exp}\left(-\frac{r_T^2}
{R_0^2(s)}\right)\right]\ ,
\label{180}
 \eeq
 where $R_0(s)=0.88\,fm\,(s_0/s)^{0.14}$ and $s_0=1000\,GeV^2$
\cite{kst2}. The energy dependent factor $\sigma_0(s)$ is defined as,
 \beq
\sigma_0(s)=\sigma^{\pi p}_{tot}(s)\,
\left(1 + \frac{3\,R^2_0(s)}{8\,\la r^2_{ch}\ra_{\pi}}
\right)\ ,
\label{190}
 \eeq
 where $\la r^2_{ch}\ra_{\pi}=0.44\pm 0.01\,fm^2$ \cite{pion} is the mean
square of the pion charge radius.  

This dipole cross section is normalized to reproduce the pion-proton total
cross section, $\la\sigma_{\bar qq}\ra_\pi=\sigma_{tot}^{\pi p}(s)$. The
saturated shape of the dipole cross section is inspired by the popular
parametrization given in Ref.~\cite{gbw1,gbw2}, which is fitted to the low-$x$ and
high $Q^2$ data for $F^p_2(x,Q^2)$ from HERA. However, that should not be used
for our purpose, since is unable to provide the correct energy dependence of
hadronic cross sections. Namely, the pion-proton cross section cannot exceed
$23\mb$. Besides,
Bjorken $x$ is not a proper variable for soft reactions, since at small $Q^2$
the value of $x$ is large even at low energies. The $s$-dependent dipole cross
section Eq.~(\ref{180}) was fitted \cite{kst2} to data for hadronic cross
sections, real photoproduction and low-$Q^2$ HERA data for the proton
structure function.  The cross section (\ref{190}) averaged with the pion wave
function squared (see below) automatically reproduces the pion-proton cross
section.

In the case of a proton beam one needs a cross section for a three-quark dipole,
$\sigma_{3q}(\vec r_1,\vec r_2,\vec r_3)$, where $\vec r_i$ are the transverse
quark separation with a condition $\vec r_1+\vec r_2+\vec r_3=0$. In order to
avoid the introduction of a new unknown phenomenological quantity, we 
express the
three-body dipole cross section via the conventional dipole cross section $\sq$
\cite{mine},
 \beq
\sigma_{3q}(\vec r_1,\vec r_2,\vec r_3) =
{1\over2}\,\Bigl[\sigma_{\bar qq}(r_1)+
\sigma_{\bar qq}(r_2)+
\sigma_{\bar qq}(r_3)\Bigr]\ .
\label{195}
 \eeq
 This form satisfies the limiting conditions, namely, turns into
$\sigma_{\bar qq}(r)$ if one of three separations is zero. Since all these
cross sections involve nonperturbative effects, this relation hardly can
be proven, but should be treated as a plausible assumption.

The 3-quark valence wave function is modelled assuming that
the dipole cross section is independent of the sharing
of the light-cone momentum among the quarks, so the wave function squared of
the valence Fock component of the proton, $\left|\Phi(\vec
r_i,\alpha_j\right|^2$ should be integrated over fractions $\alpha_i$. The
result depends only on transverse separations $\vec r_i$. The form of the
nonperturbative valence quark distribution is unknown, therefore for the
sake of simplicity we assume the Gaussian form,
 \beqn
&&\left|\Psi_N(\vec r_1,\vec r_2,\vec r_3)\right|^2 =
\int\limits_0^1 \prod\limits_{i=1}^3 d\alpha_i\,
\left|\Phi(\vec r_i,\alpha_j)\right|^2\,
\delta\left(1-\sum\limits_{j=1}^3 \alpha_j\right)
\nonumber\\ &=&
\frac{2+r_p^2/R_p^2}{(\pi\,r_p\,R_p)^2}
\exp\left(-\frac{r_1^2}{r_p^2}-\frac{r_2^2+r_3^2}{R_p^2}\right)\,
\delta(\vec r_1+\vec r_2+\vec r_3)\ ,
\label{200}
 \eeqn
 where $\vec r_i$ are the interquark transverse distances. The two scales
$r_p$ and $R_p$ characterizing the mean transverse size of a diquark and
the mean distances to the third quark.

For the sake of simplicity here we assume
that the forces binding the valence quarks are of an iso-scalar nature,
therefore the quark distribution is symmetric, i.e. $r_p=R_p$ in
(\ref{200}). In this case the mean interquark separation squared is
$\la \vec r_i^{\,2}\ra={2\over3}R_p^2 = 2\la r_{ch}^{2}\ra_p$.
See other possibilities of an asymmetric valence structure in Ref.\cite{xsect}.

Apparently, any model for the dipole cross section and valence quark distribution in the proton, must reproduce correctly data for diffractive excitation of the proton,
otherwise the Gribov corrections will come out wrong. It was demonstrated
that with the above choice of the dipole cross section and proton wave function
one reproduces quite well the results of the global analysis \cite{kklp} of single diffraction data, namely the triple-Reggeon term $\Pom\Pom\Reg$, as well as
the triple-Pomeron one $\Pom\Pom\Pom$, controlling diffractive gluon radiation.

Now we are in a position to calculate the Gribov corrections.

\subsubsection{Excitation of the valence quark skeleton of the 
proton}

The total cross sections reads,
 \beqn
\sigma_{tot}^{pA} &=&
2\int d^2b\,\left[1-
\left\la e^{-{1\over2}\,
\sigma_{3q}(r_i)\,T_A(b)}\right\ra\right] 
\nonumber \\ &\equiv&
2\int d^2b
\left[1-
\int \prod\limits_{i=1}^3 d^2r_i\,
\left|\Psi_N(r_j)\right|^2\,
e^{-{1\over2}\,
\sigma_{3q}(\vec r_k)\,T_A(b)}\right].
\label{312}
 \eeqn

Using the wave function Eq.~(\ref{200}) with $r_p=R_p$ and the cross
section (\ref{180}) we get the following forward elastic cross section,
 \beq
\left.\frac{d\sigma^{pp}_{el}}
{dp_T^2}\right|_{p_T=0} =
\frac{\gamma^2}{(1+{2\over3}\gamma)^2}\ 
\frac{\sigma_0^2(s)}{16\pi},
\label{265}
 \eeq
 where 
$\gamma = 3\la r_{ch}^2\ra_p/R_0^2(s)$.

\section{Gluon shadowing}\label{gluons}

Eikonalization of the lowest Fock state $|3q\ra$ of the proton done in
(\ref{312}) corresponds to the Bethe-Heitler regime of gluon radiation.
Indeed, gluon bremsstrahlung is responsible for the rising energy dependence
of the cross section, and in the eikonal form
(\ref{312}) one assumes that the whole spectrum of gluons is radiated in each of multiple
interactions.
However, the Landau-Pomeranchuk-Migdal effect \cite{lp,m} is known to
suppress radiation in multiple interactions. Since a substantial part of the
inelastic cross section at high energies is related to gluon radiation, the
LPM effect suppresses the cross section. This is a
quantum-mechanical interference phenomenon and it is a part of the suppression
called Gribov inelastic shadowing. In
the QCD dipole picture it come from inclusion of higher Fock states, $|3qg\ra$,
etc. Each of these states represents a colorless dipole and its elastic
amplitude on a nucleon is subject to eikonalization.

As we already mentioned, the eikonalization procedure requires the
fluctuation lifetime to be much longer than the nuclear size. Otherwise,
one has to take into account the "breathing" of the fluctuation during
propagation through a nucleus, which can be done by applying the
light-cone Green function technique \cite{kz91,kst1,kst2}. In hadronic
representation this is equivalent to saying that all the longitudinal
momenta transfers must be much smaller than the inverse mean free path of the
hadron in the nucleus. Otherwise, one should employ the path-integral technique, described above.

The c.m. energies of HERA-B, RHIC and LHC are sufficiently high to treat
the lowest Fock state containing only valence quarks as "frozen" by the
Lorentz time dilation during propagation through the nucleus.  Indeed, for
the excitations with the typical nucleon resonance masses, the coherence
length is sufficiently long compared to the nuclear size. This is why we
applied eikonalization without hesitation so far. Such an approximation,
however, never works for the higher Fock states containing gluons. Indeed,
since the gluon is a vector particle, the integration over effective mass of the
fluctuation is divergent, $dM^2/M^2$, which is the standard triple-Pomeron
behaviour. Therefore, the energy of collisions can never be sufficiently high
to neglect the large-mass tail. For this reason the inelastic shadowing
corrections, related to excitation gluonic degrees of freedom never saturates, 
and keeps rising logarithmically with energy.

There are, however non-linear effects which are expected to stop the rise
of inelastic corrections at high energies. This is related to the
phenomenon of gluon saturation \cite{glr,al} or color glass condensate
\cite{mv}. The strength of these nonlinear effect is expected to be rather mild 
due to smallness of the gluonic spots in the nucleus \cite{spots}.
The reason is simple, in spite of a sufficient
longitudinal overlap of gluon clouds originated from different nucleons,
there is insufficient overlap in the transverse plane. This fact leads to
a delay of the onset of saturation up to very high energies, since the
transverse radius squared of the gluonic clouds rise with energy very
slowly, logarithmically, with a small coefficient of the order of
$0.1\GeV^{-2}$. 

The details of the calculation of inelastic corrections related to excitation
of gluonic degrees of freedom can be found in Ref.\cite{kst2,mine}. The
numerical results
for nuclear cross sections at the energies of RHIC and LHC can be found in Ref.\cite{xsect,kps-ciofi}. As an example, the total proton-lead cross section,
calculated at $\sqrt{s}=5\TeV$ in the Glauber approximation, and corrected to Gribov shadowing related to excitation of valence quarks  and gluons, results in
$\sigma^{pPb}_{tot}=4242.5\mb;\ 4235.2\mb;$ and $4207.1\mb$ respectively.

\section{Summary}
The dipole phenomenology in QCD has been intensively developing over the past three decades, due to both theoretical efforts and precise experimental data, in particular from HERA. This theoretical tool allows to calculate the effects of Gribov inelastic shadowing
on a more solid ground and in all orders of opacity expansion. In this note we presented several explicit examples.

\vspace*{25px}

\noi
{\bf Acknowledgements:} Work was partially
supported by Fondecyt (Chile) grant 1130543.
\vspace{2ex}

\end{document}